%% file: main.tex
\documentclass{article}

\usepackage[T1]{fontenc} 
\usepackage[utf8]{inputenc} 
\usepackage{ismir,amsmath,cite,url}
\usepackage{graphicx}
\usepackage{color}
\usepackage{booktabs}
\usepackage{amsmath}
\usepackage{multirow}
\usepackage{soul}
\usepackage{booktabs}
\usepackage[labelformat=simple]{subcaption}

\usepackage{framed}
\usepackage[title]{appendix}

\usepackage{xcolor}

\title{Variable-Length Music Score Infilling via XLNet and Musically Specialized Positional Encoding}

\threeauthors
  {Chin-Jui Chang} {Research Center for IT Innovation, \\ Academia Sinica \\ {\tt reichang182@gmail.com}}
  {Chun-Yi Lee} {Department of Computer Science, \\ National Tsing Hua University \\ {\tt cylee@cs.nthu.edu.tw}}
  {Yi-Hsuan Yang} {Yating Music Team \\ Taiwan AI Labs \\ {\tt yhyang@ailabs.tw}}

\def\authorname{C.J. Chang, C.Y. Lee, and Y.H. Yang}

\usepackage[bookmarks=false,pdfauthor={\authorname},pdfsubject={\papersubject},hidelinks]{hyperref}

\begin{document}

\maketitle

\input{sections/abstract}
\input{sections/introduction}
\input{sections/related_work}

\input{sections/background}
\input{sections/methodology}
\input{sections/settings}

\input{sections/experiment}

\bibliography{ISMIRtemplate}

\newpage
\begin{appendices}
\input{sections/appendix}
\end{appendices}

\end{document}

%% file: sections/abstract.tex
\begin{abstract}
This paper proposes a new self-attention based model for \emph{music score infilling}, i.e., to generate a polyphonic music sequence that fills in the gap between given past and future contexts. 
While existing approaches 
can only fill in a short segment with a fixed number of notes, or a fixed time span between the past and future contexts,
our model can infill a variable number of notes (up to 128) for different time spans.
We achieve so with three major technical contributions. First, we adapt \emph{XLNet}, an autoregressive model originally proposed for unsupervised model pre-training, to  music score infilling.
Second, we 
propose a new, musically specialized positional encoding called \emph{relative bar encoding} 
that better informs the model of notes' position within the past and future context.
Third, to capitalize relative bar encoding, we perform \emph{look-ahead onset prediction} 
to predict the onset of a note one time step before predicting the other attributes of the note.
We compare our proposed model with two strong baselines and show that our model is superior in both objective and subjective analyses. 
\end{abstract}

%% file: sections/introduction.tex
\section{Introduction}
A growing body of research work has adopted deep learning techniques to generate music sequentially, taking only the past context as a condition while generating. 
This paper deals with a 
different
setting where both the past and future context are given, called \emph{music score infilling}
\cite{DeepBach,anticipation_rnn,latent_space_musical_inpainting,music_sketchnet,nonoto19iccc,coconet,Kosuke_Nakamura2020,esnet,infilling_piano}.
As depicted in Figure \ref{fig:music_score_inpainting}, the goal of this task is to generate a short piece of symbolic music that fits in the middle gap. The generated piece is expected to meet certain requirements, e.g., being coherent with the contexts, and having an appropriate duration span such that the generated piece will not overflow to the outside of the designated region. Models with such a capability are useful in a few scenarios. For instance, one may have an inspiration to write two segments of melodies but somehow do not know how to connect them \cite{nonoto19iccc}. Or, a music piece may have an impaired part in the middle that requires restoration~\cite{audio_inpainting}.

\input{figs/music_score_inpaint}

As reviewed in Section \ref{bg}, existing models for  score infilling can be categorized by their representation of musical scores. 
Among them, we are 
interested in the case when music is represented as a sequence of \emph{event tokens} such as note-on and note-off \cite{midi_representation}, for such a \emph{token}-based representation facilitates the use of 
the self-attention based Transformers~\cite{transformer}
for model building,
which has been shown to outperform recurrent neural networks (RNN) 
in modeling sequences~\cite{bert,karita19asru}.\footnote{Token-based representations have also been heavily adopted by recent Transformer models for sequential MIDI music generation~\cite{transformer_NADE,music_transformer, payne2019musenet, LakhNES, encode_musical_style, remi, guitar_tabs_transformer, jazz_transformer, PopMAG, cp_word}.} 
However, Transformers are originally designed for sequential generation: predicting the future given the past tokens. Adapting it to account for both the past and future contexts while generating the missing token sequence in the middle may not be trivial.

The Infilling Piano Performances (IPP)~\cite{infilling_piano} is the only pioneering work adapting Transformers for music score infilling, to our best knowledge. 
It achieves so with a simple approach of ``reordering'': the past and future contexts are concatenated and placed \emph{before} the missing segment as the \emph{prompt}. A standard Transformer decoder is then trained to autoregressively generate a continuation of a given prompt.
We note that this approach has a strong limitation: the past and future contexts need to consist of a fixed number of $c$ notes each, and the missing segment 
a \emph{fixed} number of $k$ notes 
($c$ and $k$ can be different).
Moreover, as reported in \cite{infilling_piano}, empirically their  model works well only for infilling a \emph{short} sequence with $k=16$ missing notes; for larger $k$, its performance deteriorates and the infilled sequence cannot connect well with the future context.

In \emph{text infilling}~\cite{mask_gan,zhu18arxiv,blm,inset,ilm,felix}, the music infilling counterpart in natural language processing (NLP), many approaches have been based on  Transformers. For instance, Infilling Language Model (ILM)~\cite{ilm} uses special tokens to inform the language models where to infill text. FELIX~\cite{felix} enables BERT~\cite{bert} to solve the infilling task by letting BERT predict [PAD] tokens for redundant masked positions. 
Both models can perform \emph{variable-length text infilling}, but they were both tested only on infilling short sequences with less than 10 consecutive tokens (words).


This paper proposes a new self-attention based model to  attain \emph{long} and \emph{variable-length music score infilling}. In our experiment, the model is able to generate a variable-length polyphonic infilled sequence with up to 768 tokens, or 128 notes, given the past and future contexts.\footnote{As described in Section \ref{sec:token}, we represent a musical note with 6 tokens (\texttt{PITCH}, \texttt{DURATION}, etc) in total.} 
Moreover, our model is able to infill spans of different length (e.g., 2 bars to 4 bars) without re-training 
for each span length.
We achieve so with three technical contributions. 
First, we employ \emph{XLNet}~\cite{xlnet}, a Transformer encoder-based model, 
as the model architecture for the first time 
for music generation.
Unlike other bidirectional models such as BERT\cite{bert}, XLNet can
attend to 
the past and future contexts while maintaining its autoregressive predicting order. 
We show that music infilling can be attained by XLNet via a specific factorization order of the token sequence.

Second, we point out that the original XLNet can infill only a fixed-length token sequence, because it relies on the \emph{vanilla} positional encoding\footnote{Unlike RNNs, the Transformers do not have a built-in notion of the sequential order of tokens, and thus need to rely on the so-called \emph{positional encodings} that ``assign'' positions to each token \cite{ke2021rethinking,wang2021position}. 
This is usually done by using a token's \emph{absolute} position in the sequence~\cite{transformer}, or by its \emph{relative} distance to other tokens~\cite{transformer_xl, relative_attention,liutkus2021relative}.} that requires the length of the missing segment to be known and fixed in advance.
We propose a \emph{musically specialized positional encoding} and a modification of the two-stream attention mechanism of XLNet to make it feasible for variable-length infilling. 
Specifically, our model 
represents the distance between two  tokens in terms of \emph{the number of bars} 
between them, rather than the exact number of intermediate tokens. 
Doing so, the positions of notes are also specified in a musically more meaningful way. 

Third, with our special positional encoding, we need to know the musical position of the next note to be predicted in the autoregressive generation process.
Therefore, we adapt the multi-output methodology of the \underline{C}om\underline{p}ound Word (CP) Transformer \cite{cp_word} 
to perform \emph{look-ahead onset prediction} at each timestep.
Specifically, each time, our model predicts the  \textit{content}-related tokens of the \emph{current} note (i.e., \texttt{PITCH}, \texttt{DURATION}, \texttt{VELOCITY}, and \texttt{TEMPO}), 
and the \textit{position}-related tokens (i.e., \texttt{BAR} and \texttt{SUB-BEAT}) of the \emph{next} note to look one note ahead.

For evaluation, we compare our model with the two text infilling state-of-the-arts, ILM \cite{ilm} and FELIX \cite{felix}, that we extend to accept the same token representation as ours and to generate variable-length infilled sequences. The results show that our model outperforms these two strong baselines in both objective and subjective analyses. 


For reproducibility, we open source our code at GitHub, along with examples of the infilling result.\footnote{\url{https://github.com/reichang182/variable-length-piano-infilling}}

%% file: figs/music_score_inpaint.tex
\begin{figure}[t]
    \centering
    \includegraphics[width=\linewidth]{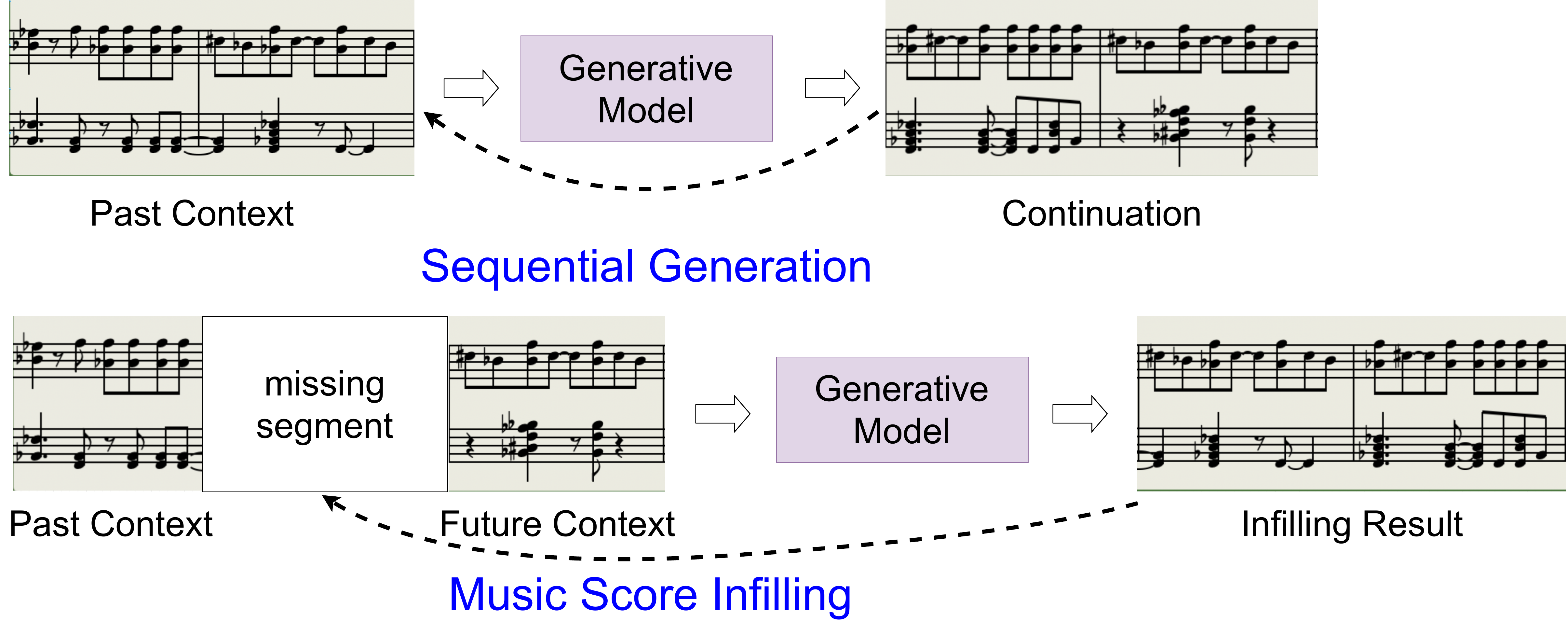}
    \caption{Comparison between sequential generation and music score infilling (a.k.a., music score inpainting).}
    \label{fig:music_score_inpainting}
\end{figure}

%% file: sections/related_work.tex
\section{Background}
\label{bg}

\input{figs/illustrate_baselines}

\subsection{Related Work on Music Score Infilling}
\label{related_work_1}

Existing work can be divided by their data representation:

\textbf{DeepBach-like}. DeepBach~\cite{DeepBach} predicts a missing note based on the information from the notes around. They use two RNNs to aggregate the past and future contexts and a feedforward neural network for the notes occurring at the same temporal position as the current target note. They use pseudo-Gibbs sampling to improve the generated score gradually. Anticipation-RNN~\cite{anticipation_rnn} introduces a constraint-RNN to the generation model, enforcing the model to consider the user-defined constraints while generating. Music InpaintNet~\cite{latent_space_musical_inpainting} also uses an RNN to integrate the information within the context. However, they use the encoder of a bar-wise variational auto-encoder (VAE) to encode measures into latent vectors first and use the decoder of the VAE to reconstruct measures from the latent vectors. Music SketchNet~\cite{music_sketchnet} extends the work of Music InpaintNet to allow their model to consider user preferences. 

DeepBach-like representation represents notes in an elegant way. But, it is mainly suitable for scores with a constant number of voices, e.g., four voices for Bach chorale and one voice for Irish and Scottish folk tunes
\cite{scottish_monophonic}, since representing multiple notes played at the same time requires more tracks added to the representation. Prior arts on DeepBach-like representation all use RNNs to deal with contexts, while we use Transformer-based models, which have been shown more powerful~\cite{museMorphose21arxiv,music_transformer}. 

\textbf{Piano roll}. Coconet~\cite{coconet} trains a convolutional neural network (CNN) to complete partial music score and uses blocked Gibbs sampling as an analogue to rewriting. Nakamura \emph{et al.}~\cite{Kosuke_Nakamura2020} use CNN with deconvolutional layers at the end. The whole model is trained under the framework of generative adversarial networks. 
ES-Net~\cite{esnet} represents music as a sequence of 
\emph{edit events}, each of which denotes either an addition or removal of a note. ES-Net is able to modify a music score while preventing the accumulation of errors that autoregressive models are prone to have.

The piano roll representation typically encodes information concerning pitch, pitch duration, and beat position only, not tempo and velocity, which are important to form an expressive piece. Moreover, as the CNN treats a piano roll as an fixed-size image, 
all existing piano roll-based models can only infill spans of fixed-length (e.g., 2 bars) once the CNN was trained. On the contrary, a single model trained with our methodology can be applied to tasks with missing spans of different length (e.g., 2 bars to 4 bars).\footnote{Moreover, both DeepBach-like and piano roll require a token to hold at each position; e.g., 100 tokens are needed to represent one second of piano performance at a temporal resolution of 10ms, regardless of how many note events there are~\cite{transformer_NADE}.} 

Aside from differences in data representation, some existing methods impose additional assumptions on data. For example, the use of bar-wise VAE in Music InpaintNet and Music SketchNet restrict their usage to cases where the missing segments start and end at precisely the start or end position of a bar, while our model is free of this restriction.  

\textbf{Event tokens}. 
IPP~\cite{infilling_piano} is the only work we are aware of that considers music as a sequence of tokens and employs the Transformer as the backbone model. 
Our work differs from theirs in two aspects. First, their model can only infill a fixed number of tokens, while ours can do variable-length infilling.
Second, while we both group tokens related to a note to a tuple (cf. Section \ref{sec:token}) \cite{transformer_NADE, cp_word}, our representation is based on the beat-based CP tokens \cite{cp_word}, which have built-in notion of bars and beats.

Converting our data to be acceptable to the aforementioned models and making them produce variable-length infilling is not trivial. 
Hence, in our experiment we adopt text infilling models \cite{ilm,felix} as the baselines, for they are both able to infill variable-length segment by design, not these music score infilling models.

\subsection{Related Work on Text Infilling}
\label{related_work_2}

\textbf{ILM}~\cite{ilm} replaces the missing segment with a single special token [BLANK]. The model learns to predict the original contents at the end of the sequence in an autoregressive manner. ILM only changes the input order of tokens and does not modify the attention mechanism of the vanilla Transformer. \textbf{FELIX}~\cite{felix} uses BERT to derive the capacity of utilizing bidirectional contexts. For variable-length infilling, the missing segment is replaced with a series of [MASK] tokens of a pre-defined length that is sufficiently long. The model learns to predict the tokens to infill and [PAD] tokens to indicate no token here. 
FELIX is different from our model and ILM in that it predicts all tokens in the missing segment at once, and does not consider previously generated tokens during generation. 
Both ILM and FELIX were tested on infilling less than 10 words in the original papers, while we consider up to 128 notes %
in our task.
Figure~\ref{fig:illustrate_baselines} illustrates how these two baselines work.

%% file: figs/illustrate_baselines.tex
\begin{figure}[t]
    \centering
    \includegraphics[width=\linewidth]{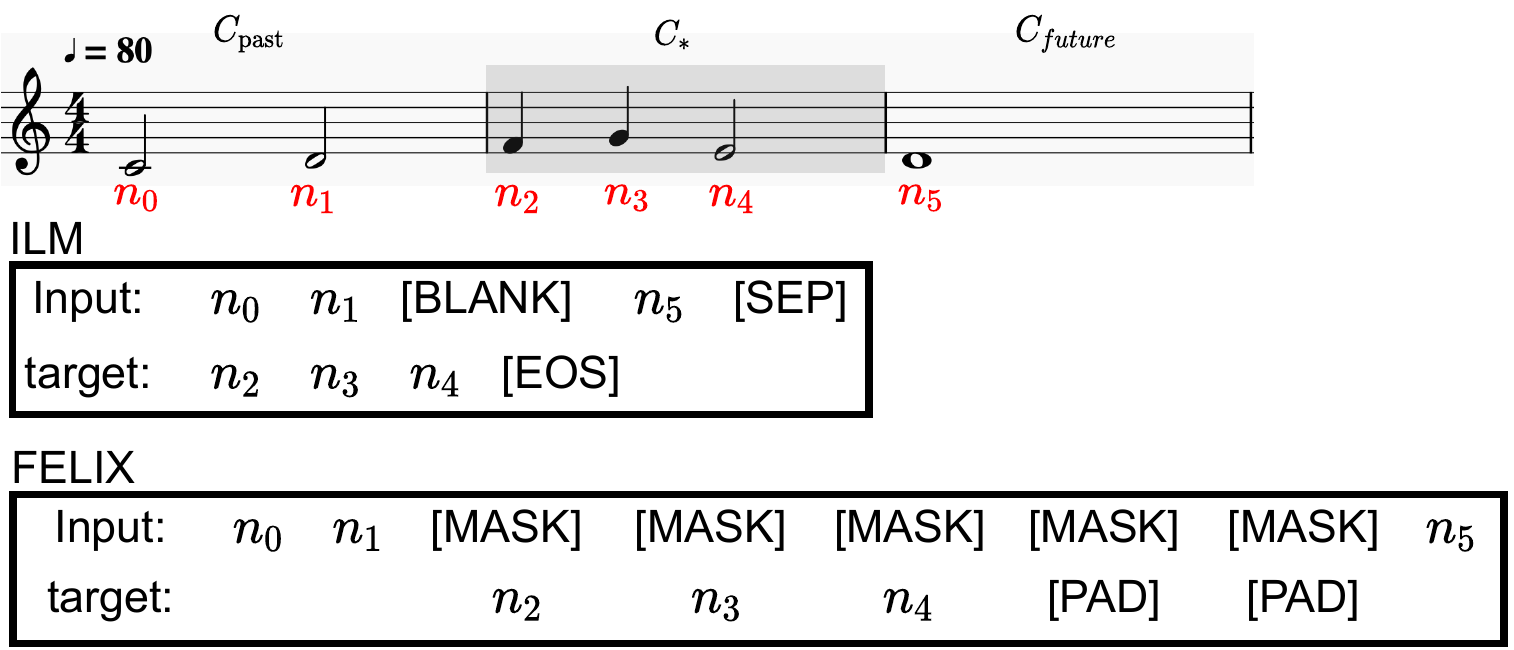}
    \caption{An illustration of how the baselines, ILM \cite{ilm} and FELIX \cite{felix}, solve the infilling task. For FELIX, the pre-defined mask length is set to five in this illustration.}
    \label{fig:illustrate_baselines}
\end{figure}

%% file: sections/methodology.tex
\section{Methodology}
\label{methodology}


We follow the definition of music score infilling in \cite{latent_space_musical_inpainting}: given a \textbf{past context} $C_\text{past}$ and a \textbf{future context} $C_\text{future}$, the task is to generate an \textbf{infilled segment} $C_*$, which connects $C_\text{past}$ and $C_\text{future}$ in a musically meaningful way. During training, the model should maximize the likelihood:
\begin{equation}
    P(C_* | C_\text{past}, C_\text{future}) \,.
    \label{eq:1}
\end{equation}
We consider in this paper the case where the model is a Transformer encoder and its input is a token sequence composing of 
$\{C_\text{past}$, a \emph{masked} version of $C_*$,  $C_\text{future}\}$. The target output is the sequence $\{C_\text{past}$, $C_*$, $C_\text{future}\}$. Model loss is computed over only the middle part related to $C_*$.

Moreover, we desire our model not to generate $C_*$ all at once, but one token at a time in an autoregressive manner. This way, the model builds $C_*$ progressively by considering its previously generated tokens. The training objective can be factorized to 
\begin{equation}
    \prod_{0 < i \leq T} P(n_i | C_\text{past}, C_\text{future}, n_{j \text{ , } 0 < j < i }) \,,
    \label{eq:2}
\end{equation}
where $n_1, ..., n_T$ are the tokens in $C_*$. Please note that, in our setting, the number of tokens $T$ is variable, 
so does the number of tokens within $C_\text{past}$ and $C_\text{future}$, respectively.

\subsection{Compound Word-based Token Representation}
\label{sec:token}

\input{figs/representation_example}
\input{tables/token_vocab}

We modify the beat-based token representation REMI~\cite{remi} and CP~\cite{cp_word} to encode our music data. 
As illustrated in Figure \ref{fig:repr_example},\footnote{For simplicity, we use monophonic pieces as examples in all the figures in the paper, though we actually use polyphonic pieces in experiment.} we describe different attributes of a musical note through six different tokens---three note-related ones, \texttt{PITCH}, \texttt{DURATION}, and \texttt{VELOCITY}, and three metric-related ones, \texttt{TEMPO}, \texttt{BAR}, and \texttt{SUB-BEAT}.\footnote{Even though it is not that reasonable to associate a \texttt{TEMPO} event with each note, 
we found that the models learn to predict similar tempos for adjacent notes, thus barely lower the quality of music.}
Table~\ref{tab:token_vocab} shows the vocabulary of the adopted token representation. 
\texttt{BAR} is encoded to 1 for encountering a new bar, while encoded to 0 for staying at the same bar. \texttt{SUB-BEAT} is the position of a note within one bar, represented in a resolution of the 16-th note. 
Following the \emph{multi-output} methodology of the CP Transformer~\cite{cp_word}, we may
consider the six tokens defining a note as a group and predict them altogether \emph{at once} at  each timestep. Accordingly, the $n_i$ in Eq. (\ref{eq:2}) is actually a ``super token'' (also referred to as a ``note'' or a CP token herefater) comprising six tokens.
%
When a CP token is masked, all its six constituent tokens are masked.

Below, we also refer to \texttt{BAR} and \texttt{SUB-BEAT} as the \emph{onset} of a note, and the other four tokens as the note's \emph{content}.


\subsection{Learning Bidirectional Contexts through XLNet}
\label{sec:learning_bidirectional_contexts}

We adapt XLNet~\cite{xlnet} to predict the masked (missing) tokens of $C_*$. 
Unlike other bidirectional models such as BERT~\cite{bert}, XLNet can effectively address Eq. (\ref{eq:2}) due to its special \textit{two-stream self-attention} mechanism.


Models such as BERT cannot address Eq. (\ref{eq:2}) because the missing parts of the input (i.e., $n_j$ in Eq. (\ref{eq:2})) is replaced with a series of masked (CP-)tokens and those parts cannot be seen by the notes to be predicted after (i.e., $n_i$, $j<i$).
The idea of two-stream self-attention is to
separate the input into two streams---the \emph{content stream} and the \emph{query stream}. 
Each masked token $n_i$ is inferred through the query stream, which masks the content of the target token at the timestep $i$. But, in inferring $n_i$, we can attend to the content of other tokens $n_j$ that are before $n_i$ through the content stream, which does not mask any tokens.

The original XLNet model is general and does not requires the masked tokens to be consecutive as the case considered in Eq. (\ref{eq:1}). 
It covers music infilling as a special case with the following specific permutation order in its \textit{permutation language modeling}: $C_\text{past} \to C_\text{future} \to C_*$.


\subsection{A New Positional Encoding}

The adapted XLNet considers both $C_\text{past}$ and $C_\text{future}$ and does well in fixed-length infilling, i.e., for scenarios where the number of tokens in $C_*$ is known or pre-defined. However, to extend the model to variable-length infilling, the vanilla positional encodings \cite{transformer_xl} employed in the original XLNet (and most Transformer-based models) to realize the sequential order of the tokens become a problem. 
\emph{Without knowing the number of tokens in $C_*$}, we cannot assign proper positional encoding to the notes in $C_\text{future}$.\footnote{Specifically, while we know the length of $C_*$ at training time, we do not know its length at inference time.} 


\input{figs/relative_bar_positional_encoding}

To address this issue, we propose a novel \emph{relative bar encoding} to replace the original
vanilla relative positional embedding \cite{transformer_xl} adopted by XLNet.
While the original positional encoding represents the relative distance between two notes in terms of the number of intermediate notes,  the proposed method represents the distance by the number of bars in between. For instance, tokens within the same bar are $0$ bar apart, and thus get $0$ for the relative-bar positional encoding. However, for notes in the next bar, the current note is one bar before, and thus will get $-1$ for the relative-bar positional encoding. 
In this way, we only need to know how many bars $C_\text{past}$ and $C_\text{future}$ are apart to assign the relative bar positional encoding to their and $C_*$'s notes.\footnote{Moreover, relative bar encoding actually  provides more musically meaningful information to models than the original positional encoding does. For example, five notes being played at the same time are considered as four notes apart for the farthest two notes by the model with the original positional encoding. However, they are actually notes with the same onset time in a score.}
See Figure~\ref{fig:rel_bar_pos_enc} for an illustration.


\subsection{Look-ahead Onset Prediction through XLNet}

\input{figs/note_information_mismatch}

While Section~\ref{sec:token} suggests that we predict the six tokens of a note $n_i$ at the same timestep, this is actually not ideal when it comes to exploiting the relative-bar positional encoding. The problem is that we need to know the onset of $n_i$ beforehand to assign a proper relative-bar positional encoding to its corresponding input (i.e., a masked CP token) to the Transformer. To this end, we propose \emph{look-ahead onset prediction}, where the onset of a note $n_i$ is inferred one timestep ahead. 
Specifically, we modify the XLNet such that \emph{the onset of $n_i$} (i.e., \texttt{BAR} or \texttt{SUB-BEAT}) is predicted along with \emph{the content of $n_{i-1}$} (i.e., four content-related tokens) at timestep $i-1$. The onset of $n_i$ is then fed to the model at timestep $i$ as the query stream input, with the content part of the query stream input masked, to infer the content of $n_i$. 
This is illustrated in Figure~\ref{fig:note_information_mismatch}(a).

There are two design details. 
First, the onset of the first note to be infilled should be provided by the user during inference phase, which may be desirable as the user can decide where to start  infilling. Second, the onset of the next note also serves as a stop signal; once the model predicts a special [EOS] token for either \texttt{BAR} or \texttt{SUB-BEAT} of the next note, the infilling process comes to an end.

\input{figs/architecture}

The overall architecture is shown in Figure~\ref{fig:architecture}.
The query stream inputs are the same as the content stream input except that the content of notes are replaced by mask tokens, since those are the parts to be inferred. The onset part of the note (i.e., \texttt{BAR} and \texttt{SUB-BEAT}) is made visible, to allow the model to exploit the onset information. We note that the \texttt{BAR} token fed as input to the query stream only tells the model whether the note is in a new bar or stays in the same bar, but not how many bars apart the current note to the other notes in $C_\text{past}$, $C_\text{future}$ and the rest of $C_*$. Therefore, the relative-bar positional encoding is still needed.

\subsection{The Necessity of Two-Stream Self-Attention}

We are now ready to elaborate more why we are in favor of XLNet instead of a Transformer decoder such as the one used by IPP~\cite{infilling_piano}. As depicted in Figure~\ref{fig:note_information_mismatch}(b) and exemplified in Figure \ref{fig:attend_problem}, to realize the look-ahead onset prediction needed by the proposed relative bar encoding, the onset-related tokens and content-related tokens in the input to a Transformer decoder would be \emph{unsynchronized}. For instance, the onset of the first input is for $n_{i+1}$, yet the content is for $n_i$. Such a mismatch impedes the Transformer decoder to attend to proper notes through the dot product of the input embeddings.
This is not a problem for XLNet, since the input tokens in either the content stream or query stream remain \emph{synchronized} (e.g., both for $n_i$). 

\input{figs/attend_problem}

%% file: figs/representation_example.tex
\begin{figure}[t]
    \centering
    \includegraphics[width=.95\linewidth]{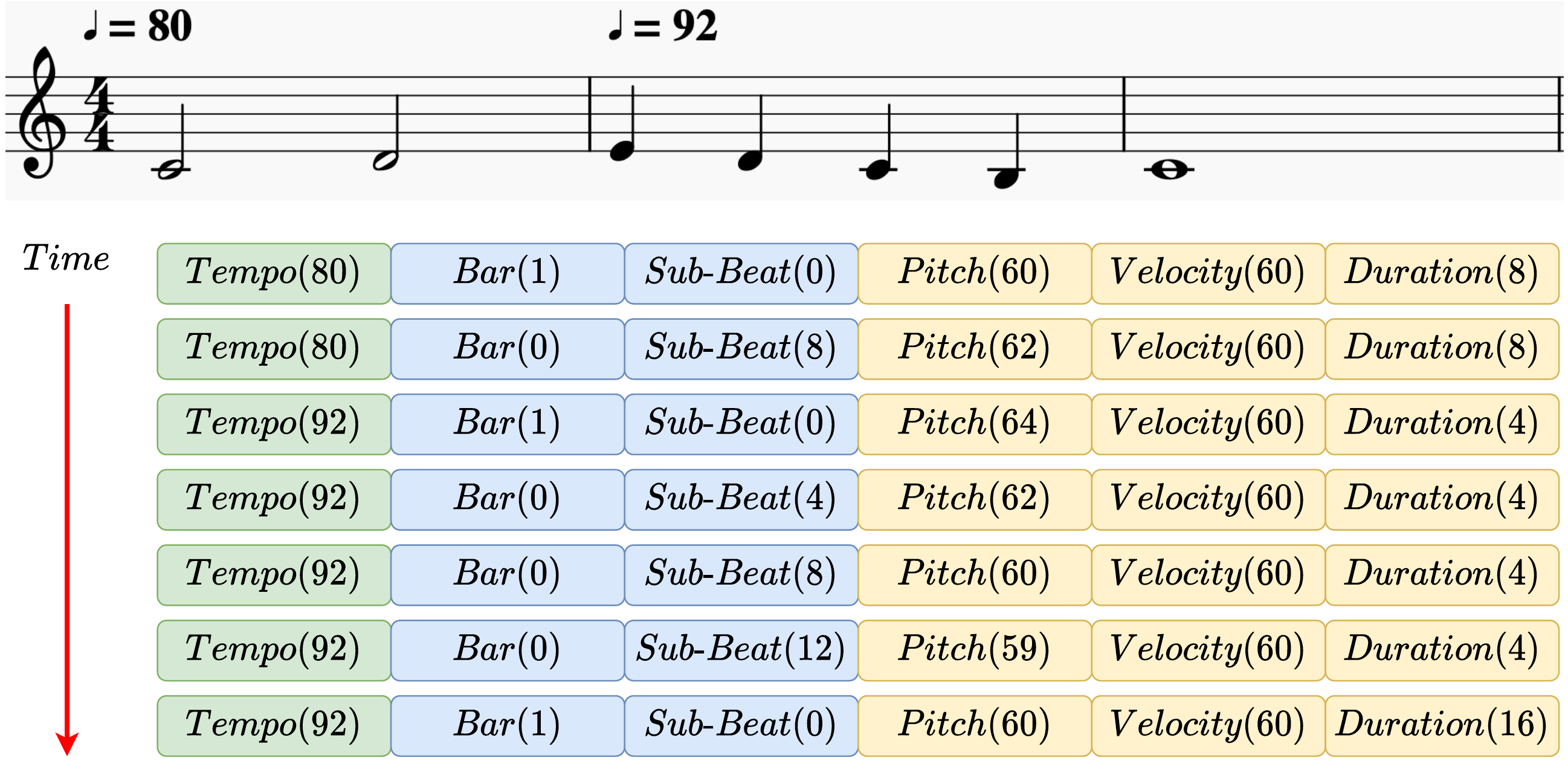}
    \caption{An example of a piece of score encoded using our representation. Note that the \texttt{BAR} and \texttt{SUB-BEAT} tokens are for positioning a note event on the time grid \cite{remi}.}
    \label{fig:repr_example}
\end{figure}

%% file: tables/token_vocab.tex
\begin{table}[t]
  \centering
  \small
      \begin{tabular}{l|c|l}
        \toprule
        Token type & Voc. size & Values\\
        \midrule
        \texttt{Tempo} & 47 & $28, 32, ..., 212$ \\
        \texttt{Bar} & 2 & $0, 1$ \\ 
        \texttt{SUB-BEAT} & 16 & $0, 1, ..., 15$ \\
        \texttt{Pitch} & 86 & $22, 23, ..., 107$ \\
        \texttt{Velocity} & 33 & $0, 4, ..., 128$ \\ 
        \texttt{Duration} & 16 & $1, 2, ..., 16$ \\
        \bottomrule
      \end{tabular}
\caption{The token vocabulary used in our experiments. Note that all the vocabulary sizes do not count special tokens such as <EOS> and <PAD>, since the use of the special tokens are model-dependent.}
\label{tab:token_vocab}
\end{table}

%% file: figs/relative_bar_positional_encoding.tex
\begin{figure}[t]
    \centering
    \includegraphics[width=\linewidth]{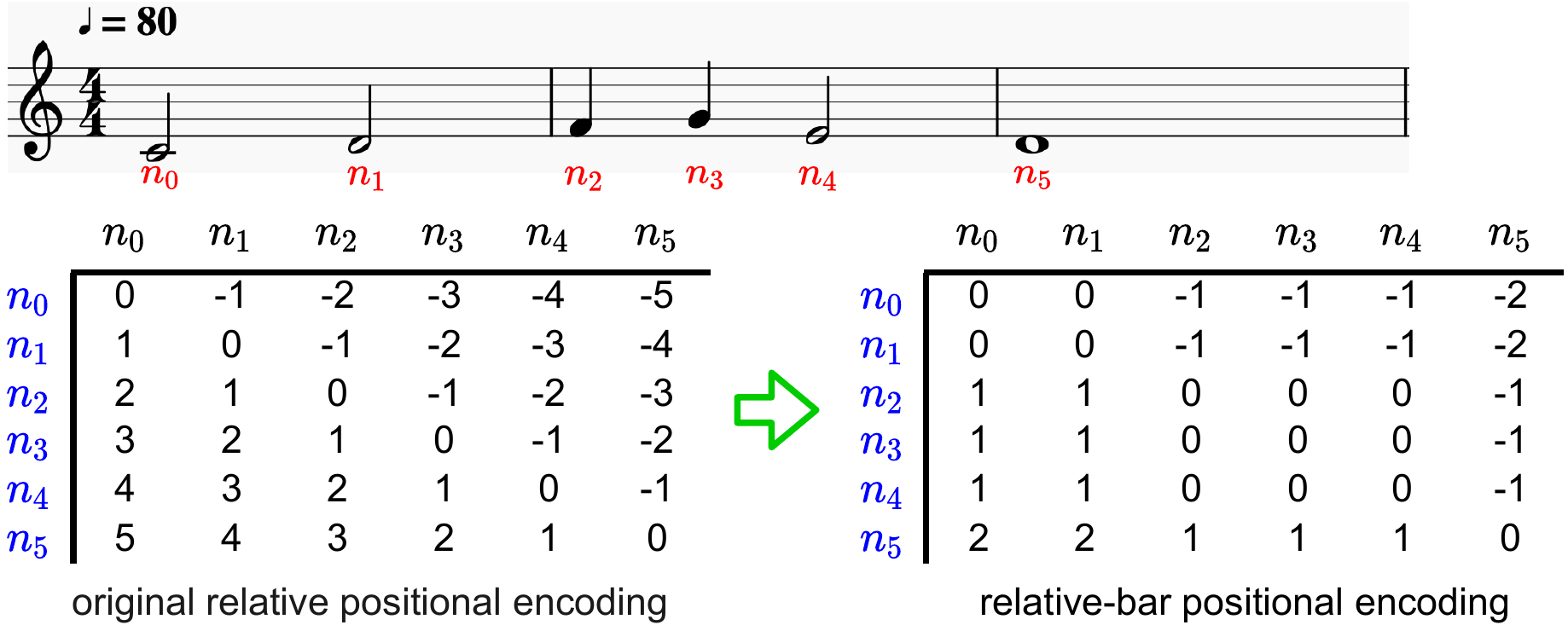}
    \caption{(Left)  the widely-used relative positional encoding vs. (right) the proposed relative-bar positional encoding. Each row shows the positional encodings of a token.}
    \label{fig:rel_bar_pos_enc}
\end{figure}

%% file: figs/note_information_mismatch.tex
\begin{figure}[t]
    \centering
    \includegraphics[width=.52\linewidth]{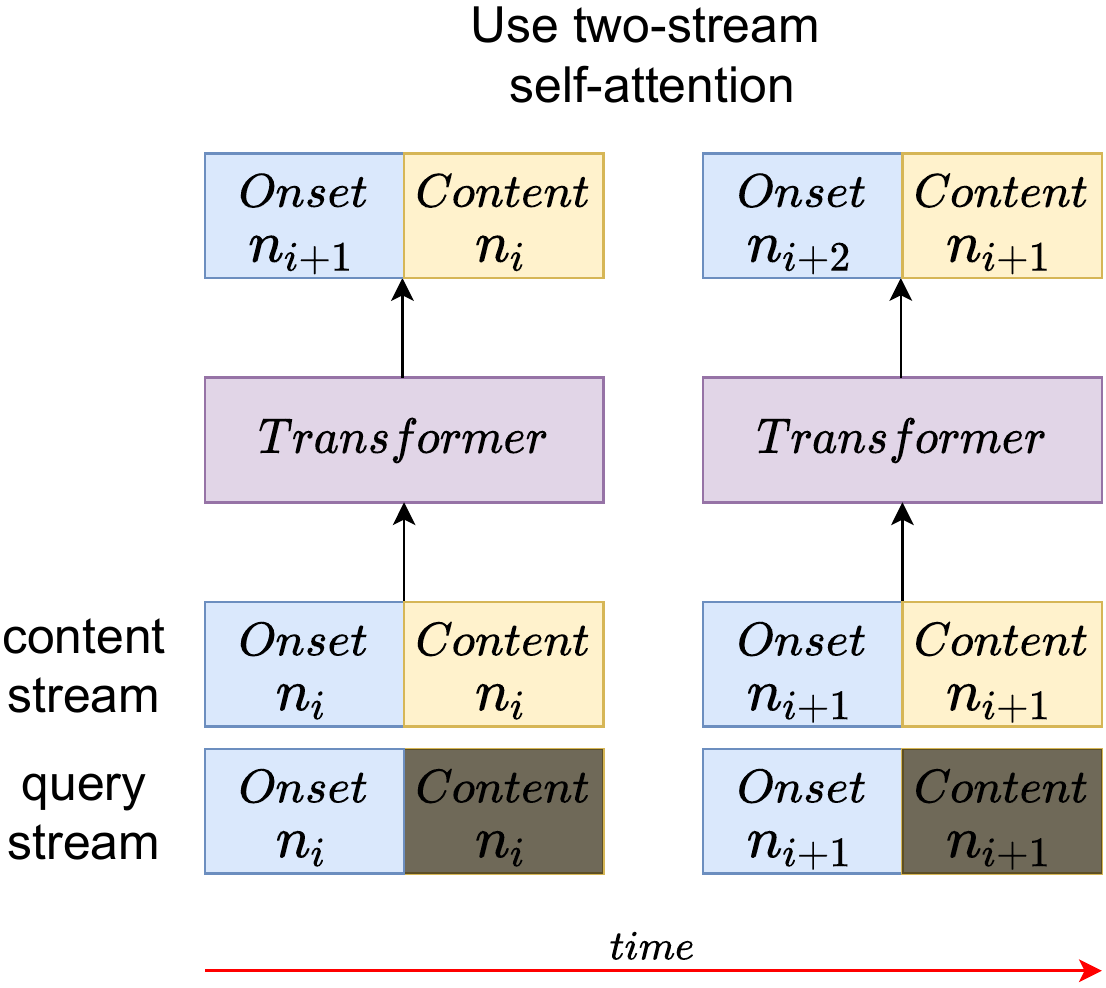} ~~~~
    \includegraphics[width=.42\linewidth]{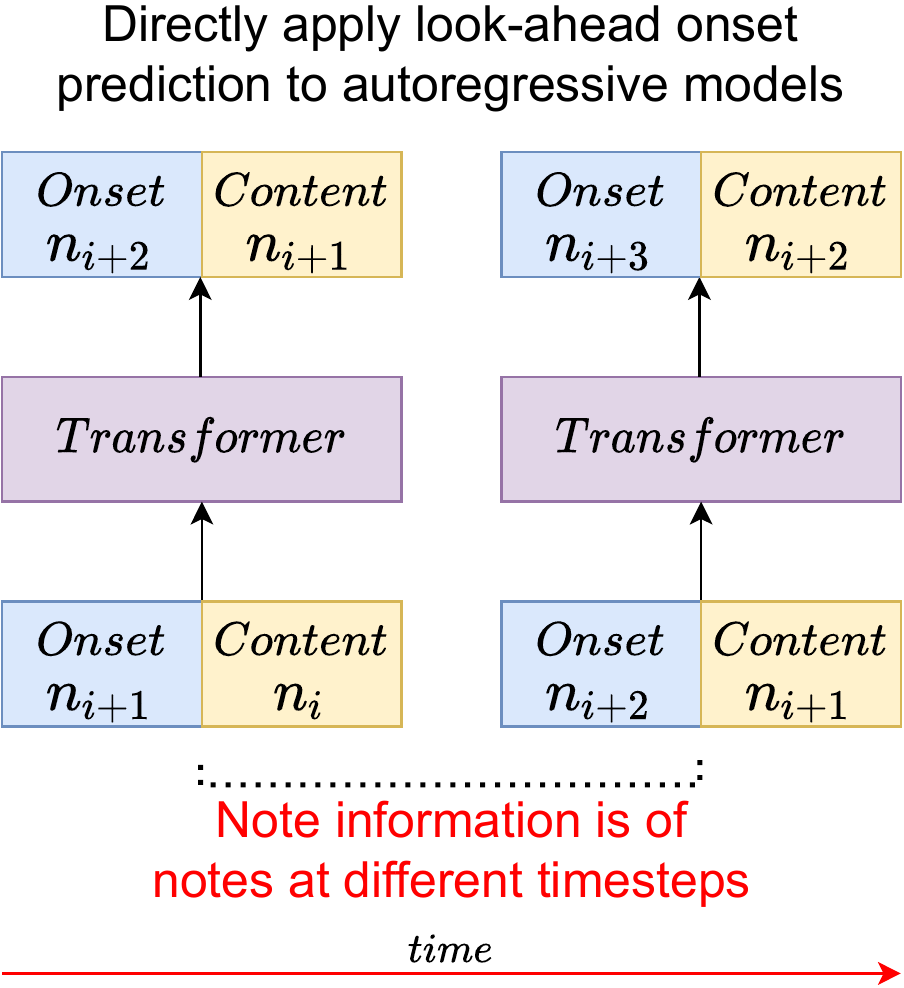} \\
    ~~~~~~(a) ~~~~~~~~~~~~~~~~~~~~~~~~~~~~~~~~~~~~~~~~ (b)
    \caption{Look-ahead onset prediction with (a) the XLNet (those shaded are masked tokens) and (b) a Transformer decoder; the input tokens in (b) are unsynchronized.}
    \label{fig:note_information_mismatch}
\end{figure}

%% file: figs/architecture.tex
\begin{figure}[t]
    \centering
    \includegraphics[width=\linewidth]{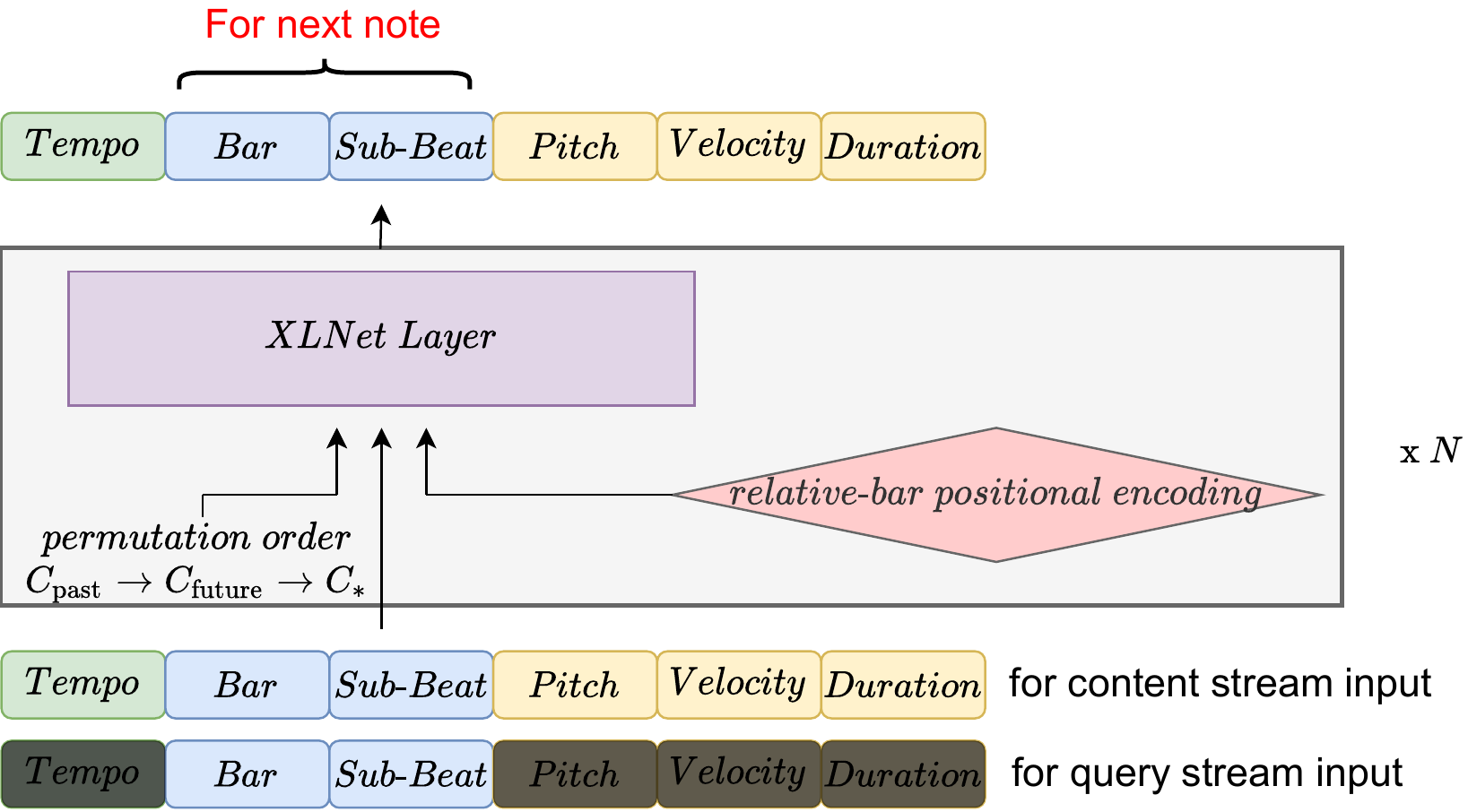}
    \caption{An illustration of the overall architecture proposed in this paper. (1) A specific permutation order of sequence is given to the model. (2) Tokens except for \texttt{BAR} and \texttt{SUB-BEAT} are masked to form the query stream input of XLNet. (3) Relative-bar positional encodings are used instead. (4) \texttt{BAR} and \texttt{SUB-BEAT} from the output of XLNet are the musical position of the next note.} 
    \label{fig:architecture}
    \vspace{-2mm}
\end{figure}

%% file: figs/attend_problem.tex
\begin{figure}[t]
    \centering
    \includegraphics[width=\linewidth]{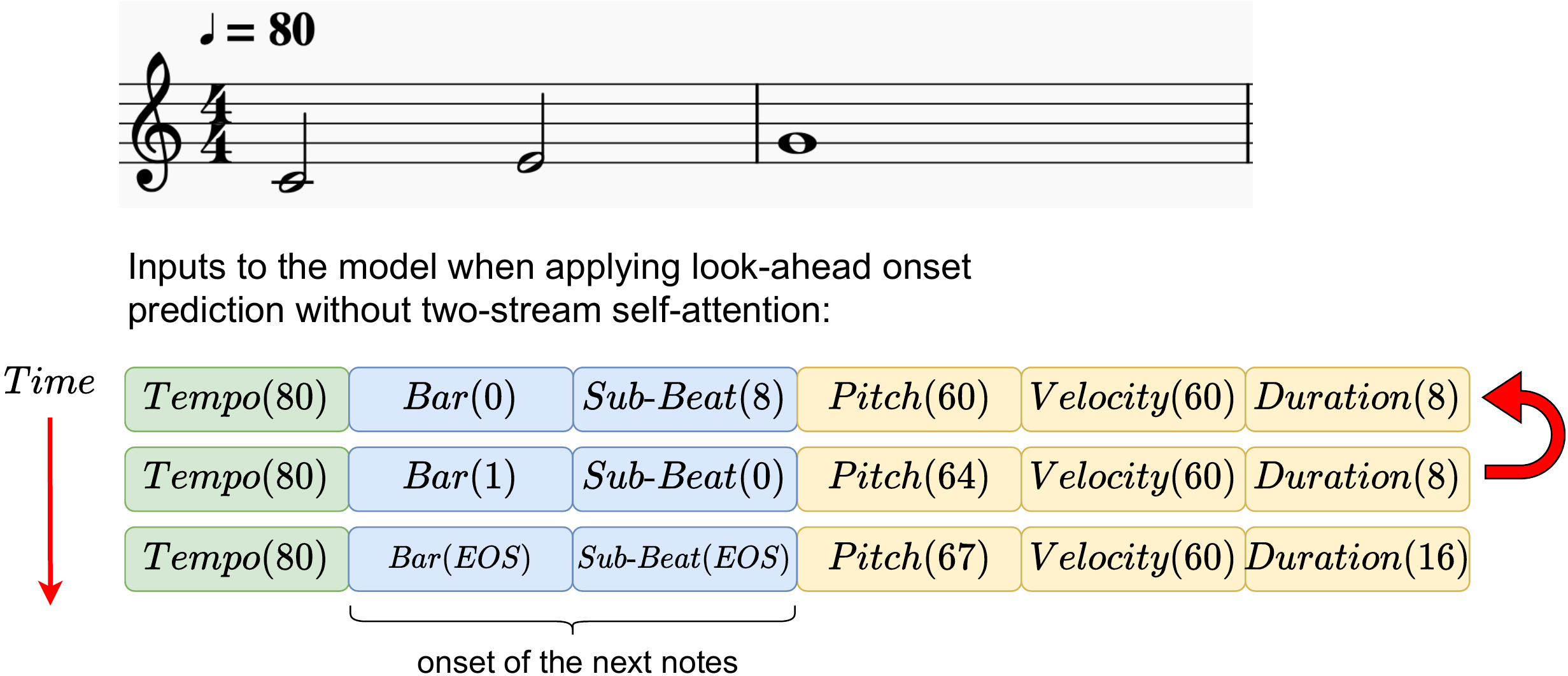}
    \caption{Illustration of the problem of a Transformer decoder such as IPP~\cite{infilling_piano} for realizing look-ahead onset prediction. To infer the third note in the sequence, it may be beneficial if the model can attend to the first note, since both notes are at the same sub-beat position (though in different bars). However, because the onset-related and content-related tokens are unsynchronized (cf. Figure \ref{fig:note_information_mismatch}(b)), the model may not be able to attend to the proper notes.}
    \label{fig:attend_problem}
    \vspace{-2mm}
\end{figure}

%% file: sections/settings.tex
\section{Experimental Setup}

\textbf{Dataset}.
We use the AILabs-Pop1k7 dataset shared publicly by Hsiao \emph{et al.} \cite{cp_word}, which contains 1,748 MIDI files of polyphonic pop piano performances, all in 4/4 time signature. We further quantize the tempo, beat position, duration, and velocity to reduce the vocabulary size, setting the 16-th note as our minimal temporal resolution for beat position and duration. 
There are on average 12.6 notes per bar.
We crop the music into 16-bar pieces with an 8-bar overlap between successive pieces, 
yielding in total 19,789 16-bar data for model training.

\textbf{Detailed Settings}.
We train all the models on 16-bar data with up to 512 CP tokens and design the experiment with the following conditions in mind. First, $C_\text{past}$ and $C_\text{future}$ must be long enough to provide sufficient contextual information. Second, the length and the musical position of $C_*$ should not be fixed, since we do not know where the missing segment starts and how many notes should be infilled for various cases in the real world. 
Consequently, for each token sequence, a range within the middle four bars, i.e., bar7 to bar10, is randomly selected to be $C_*$, such that $C_\text{past}$ and $C_\text{future}$ are at least 6 bars long, respectively. Note $C_*$ is not required to start right at the beginning of bar7. The minimum number of CP tokens to be infilled is set to half the number of CP tokens within bar7 and bar10.\footnote{And, note that we do not expect the models to rely on the absolute musical position within an entire music piece to inference, but should only rely on the note's contexts around. Thus, when providing the 16-bar data to the models, the absolute position of these 16-bar data within an entire music piece is discarded.}

A CP token is transformed before being fed to models. First, each of the six tokens composing the CP token is mapped to an embedding with size 256 using a lookup table. Then, these embeddings are concatenated and merged through a linear layer not shared across models, producing a merged embedding of size 768. All the models accept these merged embeddings as input and each has 8 heads, 12 self-attention layers, and intermediate layers of dimension 3,072. The output from these models is transformed back to probabilities with linear projection followed by softmax. At inference time, the tokens are sampled through \textit{nucleus}~\cite{nucleus}  with temperature 1.0 and threshold 0.9.

%% file: sections/experiment.tex
\section{Experimental Results}



\subsection{Objective Evaluation}

\input{figs/metric_difference}

We evaluate these models with a number of metrics proposed in \cite{jazz_transformer}, which are \emph{pitch class histogram entropy} and \emph{grooving pattern similarity}. The former provides us an indicator to the usage distribution of each pitch class within 1 bar and 4 bars, resulting in metrics $\mathcal{H}_1$ and $\mathcal{H}_4$. The latter evaluates the rhythmic pattern similarity between bars ($\mathcal{GS}$). Since the goal of the task is to connect contexts and generate fluent music, these metrics calculated on $C_*$ should be close to those calculated on $C_\text{past}$ and $C_\text{future}$. Thus we calculate the difference of metrics between $C_*$ and $C_\text{past}$, and also between $C_*$ and $C_\text{future}$. The lower the values are, the better the model is. Since these are bar-wise metrics, we let the models generate all four middle bars, i.e., bar7 to bar10, instead of a portion of them. From Figure~\ref{fig:metric_difference}, it seems that our model and ILM learn well to generate $C_*$ coherent to its contexts. The results are even close to that of the real data. In contrast, FELIX performs poorly in $\mathcal{H}_1$. It is possibly due to the dependency problems between the masked tokens, since each token is unaware of what other tokens predict, leading to uncertain tonality.

We also adapt the \emph{MIREX-like prediction test} \cite{jazz_transformer,mirex-like,janssen19ismir} 
for testing the models' ability to infer the correct answer from contexts. The test includes 1,000 questions generated from a held-out validation set, with each question comprises 6-bar $C_\text{past}$ and $C_\text{future}$. The goal of this test is to select the right infilling from four choices. A model makes a choice by calculating the average probability of the first few notes belonging to each choice, and selects the one with the highest probability. The number of notes to average from is dependent on the shortest length of all the choices. In addition, the choices could be randomly selected from a \emph{different} song, or somewhere from the \emph{same} song. 
The latter is harder since the choices are much more similar. We name them the \textbf{simple test} (choices come from different music) and the \textbf{hard test} (from the same music). The results in Table~\ref{tab:mirex} show that our model outperforms the other two consistently in both tests. It should be noted that stability is an important factor for autoregressive models in this test, since a wrongly predicted note leads to the accumulation of errors in the following prediction. While both our model and ILM could suffer from such a problem, our model still performs slightly better. 

\subsection{Subjective Evaluation}
\label{sec:evaluation-sub}

A user study is conducted with in total 30 subjects, 
where 7 of them are deemed as professionals according to a question about their musical background. Each subject is presented with 3 sets of music randomly selected from the total 15 sets of music. Within each set, a subject listens to a music piece with a missing segment, and is then presented with 4 music pieces, where 3 of them are generated by the models (Ours, ILM, and FELIX) and 1 of them is the real music without the missing segment. The subject is asked to rate each of the 4 music pieces according to its 1) \textbf{melodic fluency}: how many wrong notes are there? The fewer wrong notes, the higher the score; 2) \textbf{rhythmic fluency}: Are there notes played at the wrong time? The less, the higher the score; 3) \textbf{impression}: how much the subject is impressed? The more, the higher the score. After listening to the 4 music pieces, they are asked to choose a \textbf{favorite} one from them. From the results in Table~\ref{tab:user_study}, our model does beat the other baselines by a large margin. Our model even has a higher impression score than the real music when considering all subjects. However, there is still a gap between ours and the real data in the ``favorite'' score, suggesting that the music generated by our model is still differentiable from the real music.


\input{tables/mirex}
\input{tables/user_study}

\section{Conclusion \& Discussion}
In this paper, we have proposed a new model adapted from XLNet to address music score infilling. 
Specifically, to make the model able to perform variable-length infilling, we replace the token-based distance attention mechanism in Transformers with a musically specialized one considering relative bar distance. We have also reported evaluations showing that our model outperforms two strong baselines.

We do not pay attention to whether the past and future contexts are similar in style or theme in this work. 
It would be interesting to see in a future work whether our model can infill a nice transition between two dissimilar segments.
Moreover, 
by changing the permutation order,
our model can be applied to sequential generation in the future, 
to study whether the relative bar encoding improves the metrical structure of the generated music.

\section{Acknowledgement}

The authors are grateful to Dr. Chris Donahue, the first author of the ILM paper \cite{ilm}, for fruitful discussions during the initial phase of the project. We also thank the anonymous reviewers for their constructive suggestions, the HuggingFace team for releasing their implementation of Transformer models \cite{hugging-face-transformers} that are used in our work, and Chia-Ho Hsiung for helping with the demo website.

%% file: figs/metric_difference.tex
\begin{figure}[t]
    \centering
    \includegraphics[width=\linewidth]{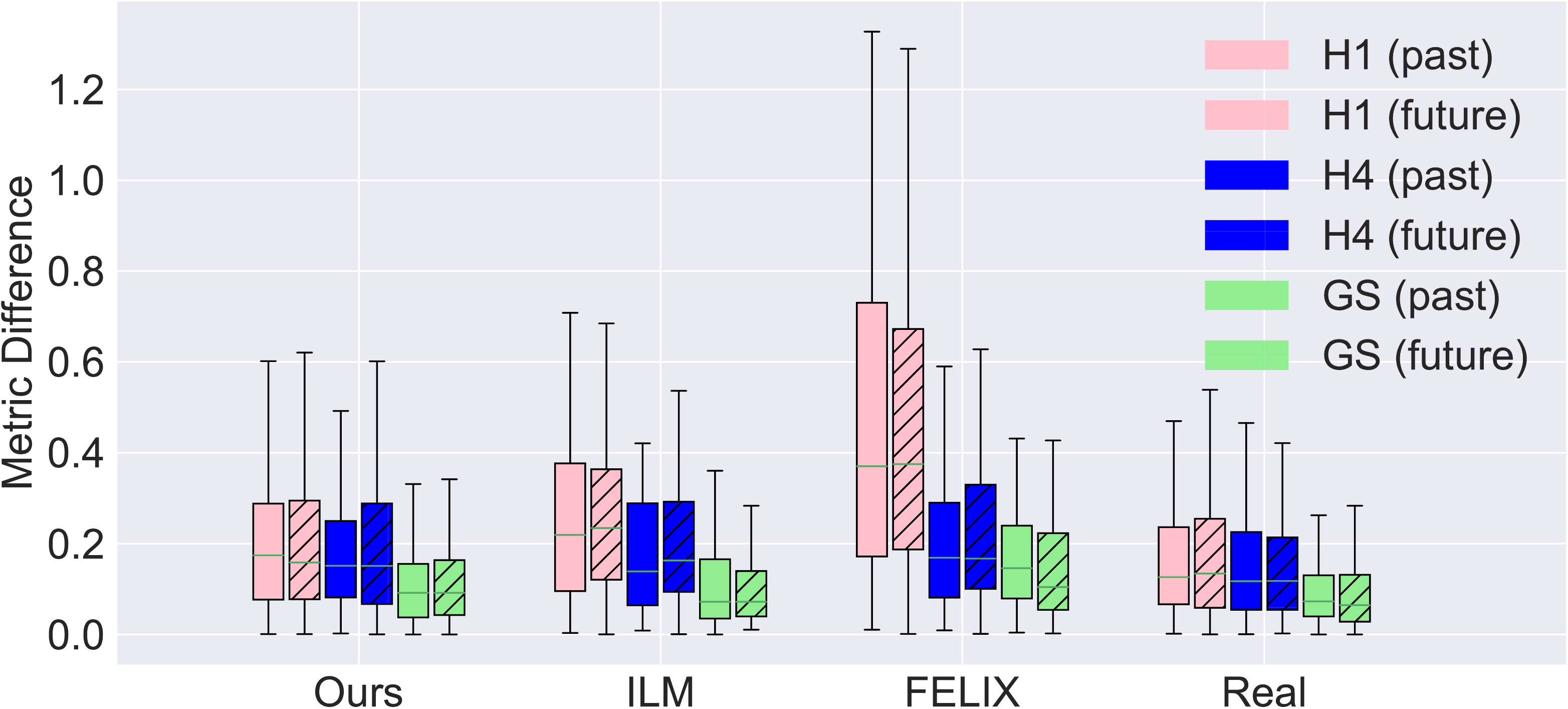}
    \caption{The difference between ``$C_*$ and $C_\text{past}$'' and between ``$C_*$ and $C_\text{future}$'' in 3 objective metrics: $\mathcal{H}_1$, $\mathcal{H}_4$, $\mathcal{GS}$.}
    \label{fig:metric_difference}
\end{figure}

%% file: tables/mirex.tex
\begin{table}[t]
    \begin{center}
     \begin{tabular}{l l l l} 
     \toprule
     & VLI (ours) & ILM \cite{ilm} & FELIX \cite{felix} \\ [0.5ex] 
     \midrule
     simple test & $\boldsymbol{0.940}^{*}$ & $0.796$ & $0.919$ \\ 
     hard test & $\boldsymbol{0.467}^{**}$ & $0.361$ & $0.398$ \\
     \bottomrule
     \multicolumn{4}{l}{\scriptsize{$^{**}$: leads all others with $p < .01$; $^{*}$: with $p < .05$}} 
    \end{tabular}
    \end{center}
    \vspace{-3mm}
    \caption{Accuracy in the MIREX-like prediction test. \textbf{VLI} denotes the proposed variable-length piano infilling model.}
    \label{tab:mirex}
\end{table}

%% file: tables/user_study.tex
\begin{table}[]
\begin{center}
\begin{tabular}{ll|ccc|c}
\toprule
                                              &       & M    & R    & I    & F    \\ \midrule
\multicolumn{1}{l|}{\multirow{4}{*}{all}} & 
VLI (ours)  & \textbf{3.27} & \textbf{3.43} & \textbf{2.93} & \textbf{27\%} \\
\multicolumn{1}{l|}{}                         & ILM \cite{ilm}  & 2.63 & 2.63 & 2.67 & 17\% \\
\multicolumn{1}{l|}{}                         & FELIX \cite{felix} & 2.83 & 2.77 & 2.57 & 16\% \\
\multicolumn{1}{l|}{}                         & Real  & \textbf{3.83} & \textbf{3.73} & \textbf{2.83} & \textbf{41\%} \\ \midrule
\multicolumn{1}{l|}{\multirow{4}{*}{pro}}     & 
VLI (ours)  & \textbf{3.08} & \textbf{3.38} & \textbf{2.77} & \textbf{24\%} \\
\multicolumn{1}{l|}{}                         & ILM \cite{ilm}  & 2.69 & 2.77 & 2.62 & 10\% \\
\multicolumn{1}{l|}{}                         & FELIX \cite{felix} & 2.85 & 2.62 & 2.62 & 14\% \\
\multicolumn{1}{l|}{}                         & Real  & \textbf{3.77} & \textbf{3.92} & \textbf{2.92} & \textbf{52\%} \\ \bottomrule
\end{tabular}
\end{center}
\vspace{-3mm}
\caption{Results of the user study: mean opinion scores in 1--5 in \textbf{M} (melodic fluency), \textbf{R} (rhythmic fluency), \textbf{I} (impression), and percentage of votes in \textbf{F} (favorite), from `all' the participants or only the music `pro'-fessionals.}
\label{tab:user_study}
\end{table}

%% file: sections/appendix.tex
\section{Appendices}
\subsection{Two-Stream Self-Attention in XLNet}
The two-stream self-attention mechanism of XLNet~\cite{xlnet} is implemented using two sets of hidden representations, i.e., the content stream $\mathbf{h}$ and the query stream $\mathbf{g}$. The content stream $\mathbf{h}$ is similar to the standard hidden states in Transformers:
\[\mathbf{h}^l_t = \text{Attention}(\mathbf{Q} = \mathbf{h}^{l-1}_t, \mathbf{KV} = \mathbf{h}^{l-1}_{\leq t}),  \Big(\mathbf{h}^0_t = e(x_t)\Big)\,,\] 
where $l$ is the layer index, $t$ is the time step, and $e(x_t)$ is the embedding of the input $x_t$. Calculations of the query ($\mathbf{Q}$), key ($\mathbf{K}$), and value ($\mathbf{V}$) in an attention operation~\cite{transformer} of the the content stream all depend on the content stream itself. Since $\mathbf{h}^{l}_t$ can attend to $\mathbf{KV} = \mathbf{h}^{l-1}_t$, $\mathbf{h}^l_t$ is able to see $\mathbf{x}_t$ by attending to $\mathbf{h}_t$ in each preceding layers.

In contrast, elements in the query stream $\mathbf{g}_t$ is designed to not see $x_t$:
\[\mathbf{g}^l_t = \text{Attention}(\mathbf{Q} = \mathbf{g}^{l-1}_t, \mathbf{KV} = \mathbf{h}^{l-1}_{<t}), \Big(\mathbf{g}^0_t = \mathbf{w}\Big)\,,\]
where $\mathbf{w}$ is a trainable vector. Even though $\mathbf{g}$ also uses the content stream to calculate $\mathbf{K}$ and $\mathbf{V}$, $x_t$ is invisible to $\mathbf{g}^l_t$ since $\mathbf{g}^{l}_t$ can only attend to $\mathbf{KV} = \mathbf{h}^{l-1}_{<t}$ and not $\mathbf{h}^{l-1}_t$. 

The content stream only serves as the content for the query stream to attend to, while the output $\mathbf{g}^L_t$ from the last layer of the query stream (not the content stream) is used for predicting the output corresponding to $x_t$.

\subsection{Implementation of Baseline Models}
We refer to the original papers and implement both the two baselines ILM \cite{ilm} and FELIX \cite{felix} based on the codebase of the Hugging Face Transformers~\cite{hugging-face-transformers} on our own. 
Specifically, we use Hugging Face's implementation of Transformer-XL~\cite{transformer_xl} for ILM, and BERT~\cite{bert} for FELIX. We set the number of layers, the number of heads, and the embedding size to be the same for our model and the baselines for a fair comparison. Our implementation of the two baselines is also provided in our GitHub repository.

\subsection{Training Loss}
The training loss curves associated with different models are shown in Figure~\ref{fig:training_loss}. We found that FELIX requires a lower learning rate to learn stably, leading to the slower descending of the curve. Nonetheless, all the models are able to decrease the training loss to reasonably low, i.e., 0.09 for ours, 0.12 for ILM, and 0.23 for FELIX.

\input{figs/training_loss}

\subsection{Controllability of the Number of Infilled Bars}
There is another benefit of using the relative-bar encoding proposed in this paper. The number of bars we want our model to generate is specified by the user as the distance between the past context $C_\text{past}$ and the future context $C_\text{future}$. After training our model on data with missing segments ranging from 1 bar to 4 bars, our model can infill a piece of originally 12-bar music ($C_\text{past}$ and $C_\text{future}$ are of 6 bars, respectively) to have 13 to 16 bars. In contrast, the number of bars that the two baselines are to infill are decided by the models themselves, because both of them do not take in the time span (or, number of bars) between $C_\text{past}$ and $C_\text{future}$ as input and the user is not able to provide such signals to the models. 
ILM autonomously stops infilling when it generates the [EOS] token, while FELIX stops infilling when it generates a [PAD] token.

In our demo website (link available on our GitHub repo; or visit \url{https://jackyhsiung.github.io/piano-infilling-demo/}), we provide examples of the infilling result of our model and the two baselines when $C_\text{past}$ and $C_\text{future}$ are four bars apart. 
While the proposed model can indeed generate segments $C_*$ with four bars long as expected, both  ILM and FELIX tend to generate much shorter segments. 
In our subjective evaluation (cf. Section \ref{sec:evaluation-sub}), we directly concatenate the infilled segments generated by ILM and FELIX with the future contexts, instead of padding zeros to make the infilled segments four bar long (because this would lead to unnatural break between the infilled segments and the future contexts).

We also show in our demo website examples of the infilling result of our model when exactly the same previous and future contexts $C_\text{past}$ and $C_\text{future}$ are given, but $C_\text{past}$ and $C_\text{future}$ are either 1, 2, 3, or 4 bars apart. 
This demonstrates the ability of our model to infill segments with different time spans. We do not evaluate this part in the user study, for neither ILM nor FELIX has this functionality.

%% file: figs/training_loss.tex
\begin{figure}[t]
    \centering
    \includegraphics[width=\linewidth]{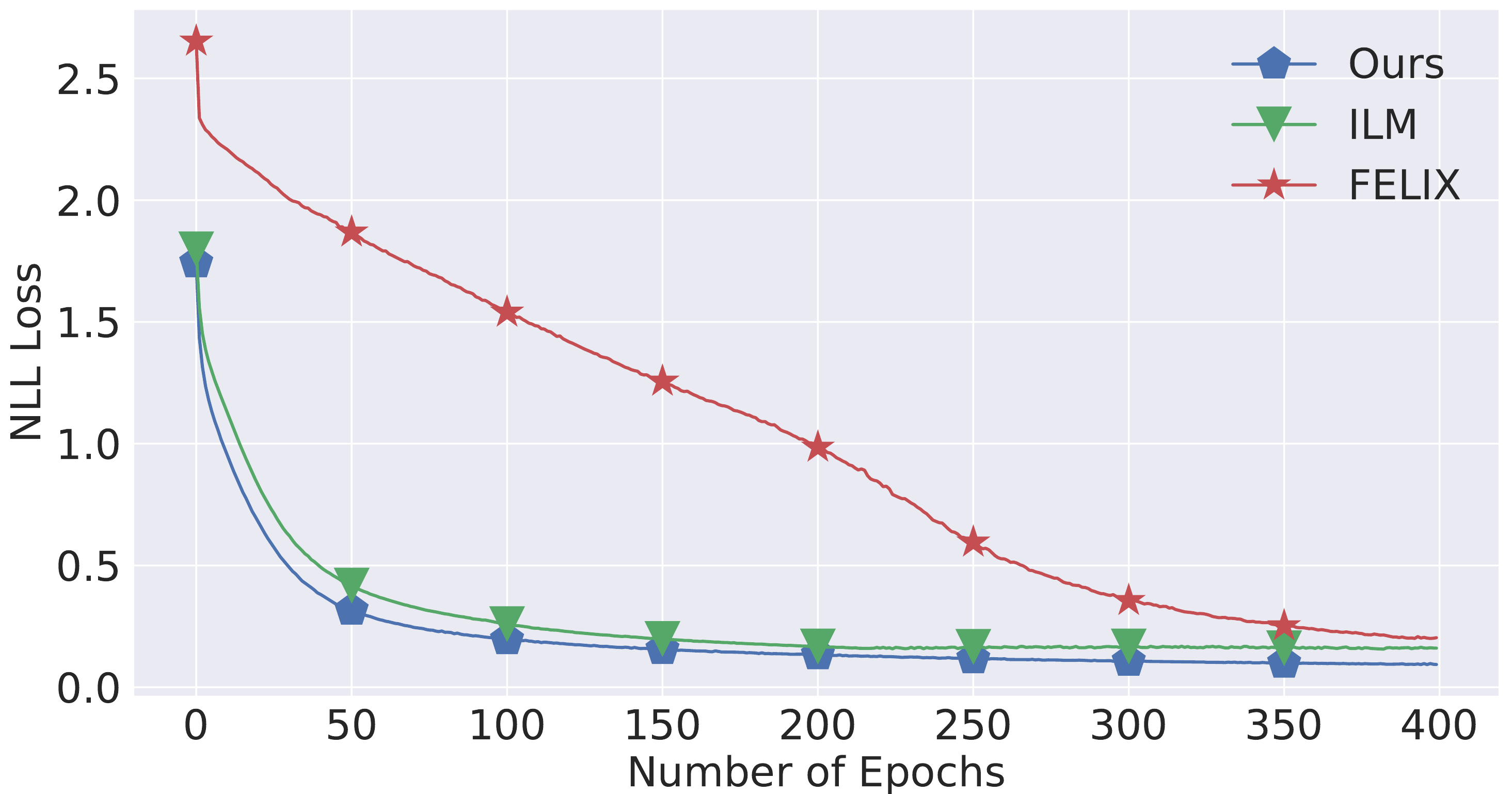}
    \caption{The evolution of training negative log likelihood (NLL) loss as a function of the number of training epochs.}
    \label{fig:training_loss}
\end{figure}